\documentclass[twocolumn,showpacs,preprintnumbers,amsmath,amssymb]{revtex4}
\usepackage[utf8]{inputenc}

\usepackage{graphicx}
\usepackage{psfrag}

\begin{document}

\preprint{TUD-ITP-TQO/04-2009-V091222}

\title{A Bose gas in a single-beam optical dipole trap}

\author{Lena Simon}
	\affiliation{Institut f\"{u}r Theoretische Physik, 
Technische Universit\"at Dresden, D-01062 Dresden, Germany}
\author{Walter T. Strunz}
\affiliation{Institut f\"{u}r Theoretische Physik, 
Technische Universit\"at Dresden, D-01062 Dresden, Germany}

\date{\today}

\begin{abstract}
We study an ultracold Bose gas in an optical dipole trap consisting of 
one single focused laser beam. An analytical expression for the 
corresponding density of states beyond the usual harmonic approximation 
is obtained. We are thus able to discuss the existence of a critical 
temperature for Bose-Einstein condensation and find that the phase
transition must be enabled by a cutoff near the threshold. Moreover, we study 
the dynamics of evaporative cooling and observe significant deviations 
from the findings for the well-established harmonic approximation. 
Furthermore, we investigate Bose-Einstein condensates in such a trap 
in Thomas-Fermi approximation and determine analytical expressions for 
chemical potential, internal energy and Thomas-Fermi radii beyond the 
usual harmonic approximation. 
\end{abstract}

\maketitle
\section{Introduction}
Since the first observation of Bose-Einstein condensation in dilute 
atomic vapors in a remarkable series of experiments 
\cite{And95,Dav95,Brad95}, ultracold quantum gases 
represent an immensely active field of research for experimental 
scientists and theorists alike. These systems offer an
unprecedented variety of possibilities to manipulate and 
investigate many-body quantum phenomena \cite{Pet02,Pit03,Blo08}.

A common approach to reach the Bose-Einstein condensation phase 
transition involves laser cooling of the atomic vapor after which atoms 
in weak-field seeking states are transferred into a magnetic 
trap \cite{Cor02,Kett02}, followed by evaporative cooling. 
While for magnetic traps the potential depends on the magnetic 
substate of the atoms, optical dipole traps have
been used to store atoms independently of their magnetic substate. To give a recent example, 
Bose-Einstein condensation of Strontium has been achieved in 
crossed laser beams \cite{Ste09,Esc09}. 

A focused laser beam provides an almost perfect realization of a 
conservative trapping potential for neutral atoms \cite{Tak96}, and is the basis for the
growing field of ultracold gases in optical lattices \cite{Frie98}. 
The first attempts to obtain quantum degeneracy in an optical dipole 
trap were made by Chu et. al. \cite{Ada95}, who employed two crossed 
laser beams. Multiple implementations of optical and 
evaporative cooling techniques may be used, which allow phase space 
densities close to quantum degeneracy \cite{Ker99,Han00,Ido00}. 
Chapman et al. managed to condense $^{87}\mathrm{Rb}$ in a dipole 
trap consisting of two crossed $\mathrm{CO}_2$-laser beams by means 
of evaporative cooling \cite{Bar01} and all-optical crossed-beam setups have 
also enabled the first observation of Bose-Einstein condensation of 
Cesium and Ytterbium atoms \cite{Web03,Tak03}. 
Optical traps consisting of one single focused laser beam were used 
to produce degenerate Fermi gases \cite{Gran02}, Fermi gas mixtures 
\cite{Luo06}, to achieve condensation of $^{87}\mathrm{Rb}$ \cite{Cen03}, 
and they are a popular way to trap and store cold gases 
for multiple needs \cite{Kae09,Szc09,Yan07,Tho08,Fan09,Bal09}.  

Usually the potential of such an optical trap is approximated 
harmonically for low average temperatures of the atoms compared to the 
trap depth \cite{Gran02,Luo06}. For some applications, however,
e.g. for evaporative cooling, typical temperatures need not be that low. 
Our work is motivated by an experiment of Helm and 
coworkers \cite{Kae09} who use a single focused $\mathrm{CO}_2$-laser 
beam to cool and store an ultracold cloud of 
$^{87}\mathrm{Rb}$-atoms. We aim at a more detailed understanding
of the evaporation dynamics and of the properties of the ultracold gas
in a single-beam trap.
Crucially, our analysis is based on the true dipole trap potential
without invoking the usual harmonic approximation. 

Our article is organized as follows. In section \ref{secdensityofstates} 
we manage to derive the density of states of the dipole trap 
based on the usual semiclassical expression without invoking further
approximations. For low energies we recover the well-known expression for
the harmonically approximated potential; for energies near threshold, 
however, we uncover a singularity. Subsequently, we will discuss the 
critical temperature for Bose-Einstein condensation in such a trap in 
section \ref{seccriticaltemp}. Remarkably,
without further thought the singularity 
in the density of states prevents Bose-Einstein condensation from
occurring. However, a cutoff -- most likely introduced by gravity -- 
removes this singularity and paves the way for Bose-Einstein 
condensation. In section \ref{secevcooling} the 
influence of the modified density of states on
evaporative cooling will be studied and significant differences
to the usual harmonic approximation will be revealed. 
Finally, in section \ref{secbec} we investigate the
condensate wave function in Thomas-Fermi approximation in such single-beam 
optical dipole traps. We conclude with final remarks in 
section \ref{secconclusion}.

\section{Density of states}
\label{secdensityofstates}
We discuss an optical dipole trap consisting of a single focused
Gaussian far detuned laser beam. The potential that affects atoms 
in such a trap is given by \cite{Mey99} 

\begin{equation}
U_{\rm dip}(\rho,z)=-\frac{U_0}{\left(1+z^2/z_0^2\right)}
\exp\left(\frac{-2\rho^2}{w_0^2\left(1+z^2/z_0^2\right)}\right)
\label{eqdipolpot}
\end{equation} 

which displays an azimuthal symmetry along the propagation ($z$-) axis.
The energy $U_0>0$ is the maximum depth of the trap at $\rho=0, z = 0$. 
Here, $z_0$ denotes the so called Rayleigh length and $w_0$ the beam 
waist of the laser which is connected to the laser wave length through
$w_0=\sqrt{\lambda z_0 / \pi}$. 

\begin{figure}[htb]
\includegraphics[width=32mm]{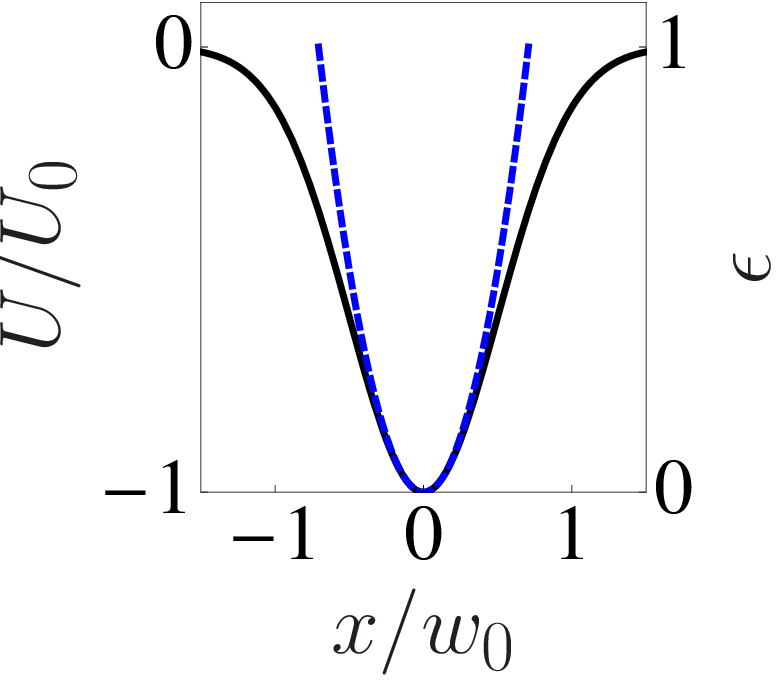}
\includegraphics[width=53mm]{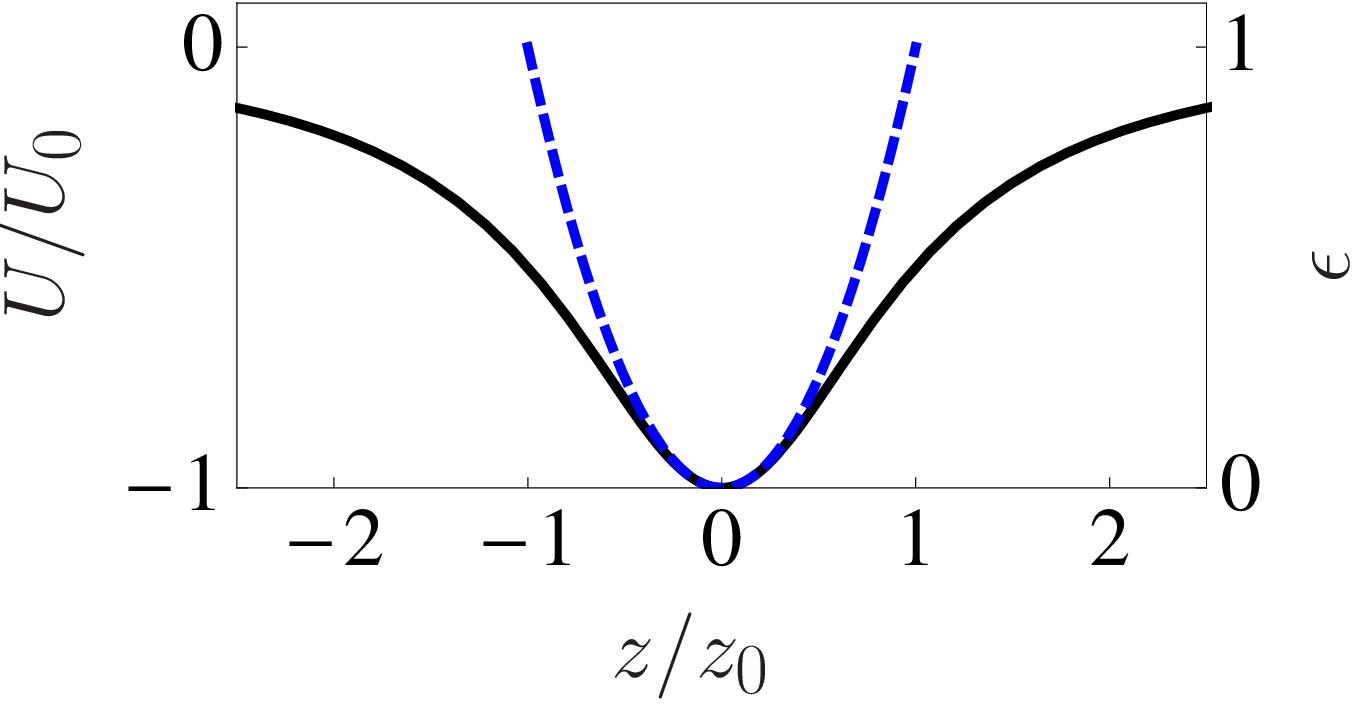}
\caption{Dipole trap potential (black, solid) and its harmonic
approximation (blue, dashed), at $z=y=0$ (left) and $x=y=0$ (right). }
\label{figpotentiale}
\end{figure}

If the average thermal energy $k_{\rm B}T$ of the atoms in the trap is much 
lower than $U_0$, the potential can be approximated harmonically 
which leads to the expression
\begin{equation}\label{eqharmonic}
U_{\rm ha}(\rho,z)=U_0\left(-1+\frac{2\rho^2}{w_0^2}+\frac{z^2}{z_0^2}\right) \ .
\end{equation}
This harmonic approximation is usually employed to 
describe cold atoms in
dipole traps \cite{Gran02,Luo06}.

For a proper discussion of the thermodynamic properties of a Bose
gas we aim at the true density of states $g(E)$ \cite{Pet02} of the 
dipole trap; the number of single-particle eigenstates between $E$ 
and $E+dE$ is then given by $g(E)dE$. In the usual semiclassical 
approximation the phase space is divided into Planck cells of 
size $(2\pi\hbar)^3$ 
such that the density of states becomes
\begin{eqnarray}\label{eqsemiclassic}
g(E)&=&\frac{1}{(2\pi\hbar)^3}\int d^3r d^3p \ 
\delta\left(E-U({\bf r})-\frac{{\bf p}^2}{2m}\right)
\\ \nonumber
&=&\frac{2\pi(2m)^{3/2}}{(2\pi\hbar)^3}\int \limits_{U({\bf r}) 
\leq E} d^3r\sqrt{E-U({\bf r})} \ .
\end{eqnarray}

In the following it turns out to be convenient to work with a
scaled energy (which is zero at the trap center and one at the threshold)
and the accompanying scaled density of states:
\begin{equation}
\epsilon = \frac{E+U_0}{U_0};\;\;g(\epsilon) = U_0 g(E),
\end{equation}
such that $g(\epsilon)d\epsilon = g(E)dE$.

Remarkably, inserting the full dipole trap potential (\ref{eqdipolpot}) in the
semiclassical formula (\ref{eqsemiclassic}) allows us to
express the density of states in terms of known special functions.
We find
\begin{eqnarray}
g_{\rm dip}(\epsilon)= G\cdot
\sqrt{\epsilon}  \Bigg\{5F\left(\arctan\left(\sqrt{\frac{\epsilon}{1-\epsilon}}\right),\frac{1}{\epsilon}\right)
\nonumber
\\
- \left(11-\frac{1}{1-\epsilon }\right)E\left(\arctan\left(\sqrt{\frac{\epsilon}{1-\epsilon}}\right),\frac{1}{\epsilon}\right)\Bigg\} \ .
\label{eqzustandsdichtedipolfalle}
\end{eqnarray}

\begin{figure}[htb]
\includegraphics[width=75mm]{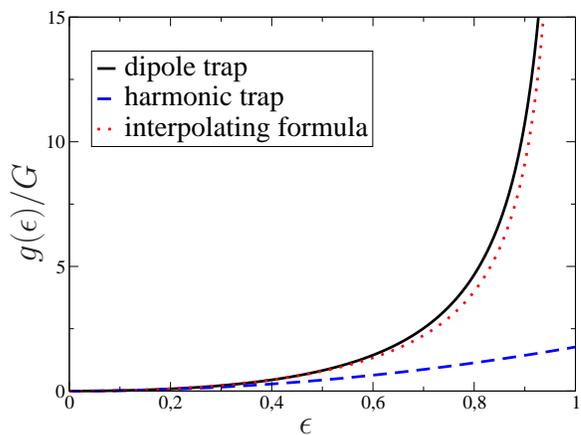}
\caption{Comparison of the densities of states of the dipole trap potential (black, solid line) and its harmonic approximation (blue, dashed line),
along with the simple approximate formula (\ref{simpleformula}) (dotted line).}
\label{figzustandsdichte}
\end{figure}

Here, $G = \frac{4\pi^2(2m U_0)^{3/2}}{9(2\pi\hbar)^3}z_0w_0^2$ is a
dimensionless constant (see below for a more convenient expression)
and $F(\phi,m)$ and $E(\phi,m)$ denote the elliptic integrals \cite{Abr64} 
of the first and second kind, respectively, with
\begin{eqnarray}
F(\phi,m)&=&\int_0^{\phi}(1-m\sin^2\theta)^{-1/2}d\theta
\nonumber
\\
E(\phi,m)&=&\int_0^{\phi}(1-m\sin^2\theta)^{1/2}d\theta \qquad \
\nonumber
\\
\nonumber
\\
&\text{for}&-\pi/2 <\phi < \pi/2 \ .
\nonumber
\end{eqnarray}
Expanding expression (\ref{eqzustandsdichtedipolfalle}) up to second order 
for small energies leads to the well-known density of states of the 
harmonic oscillator potential (\ref{eqharmonic}), as expected. We find
\begin{equation}
\label{harmonicapprox}
\lim_{\epsilon \to 0}g_{\rm dip}(\epsilon) = g_{\rm ha}(\epsilon)=
G\cdot\frac{9\pi}{16}\epsilon^2=\frac{U_0^3}{2(\hbar \bar{\omega})^3}\epsilon^2 \ , 
\end{equation} 
with the geometric mean $\bar{\omega}$ of the trap frequencies
derived from the harmonic approximation (\ref{eqharmonic}).
Thus, the prefactor in (\ref{eqzustandsdichtedipolfalle})
can also be written as $G=\frac{8}{9\pi}\frac{U_0^3}{(\hbar \bar{\omega})^3}$
which is typically a large number. The harmonic approximation
(\ref{harmonicapprox})
agrees fairly well with the true
density of states of the dipole trap potential $g_{\rm dip}(\epsilon)$ for 
energies $\epsilon$ up to about one third of the potential depth as can be
seen in Figure \ref{figzustandsdichte}. Clearly, for larger energies
the two density of states differ significantly.
Expression (\ref{eqzustandsdichtedipolfalle}) displays a $\frac{1}{1-\epsilon}$-singularity (and an additional logarithmic singularity from $F(\phi,m)$) as $\epsilon \to 1$. As a simple interpolating
formula we found
\begin{equation}\label{simpleformula}
 g_{dip}(\epsilon)\approx G\left(\frac{9\pi}{16}\epsilon^2 + \frac{\epsilon^{5/2}}{1-\epsilon}\right) \ ,
\end{equation}
which has correct limits as $\epsilon \to 0$ and $\epsilon \to 1$
(see Fig. \ref{figzustandsdichte}). 

The singularity at the potential edge has profound implications for the
thermodynamic properties of a Bose gas in this trap. In particular, we next
want to investigate its
influence on the critical temperature $T_{\rm C}$ for Bose-Einstein 
condensation.

\section{Critical Temperature}
\label{seccriticaltemp}

Bose-Einstein condensation occurs once the total number of atoms
exceeds the maximum possible number of atoms in excited states which is set
by the Bose distribution $f(\epsilon)$ with zero chemical potential.
In other words, using our scaled units, the
critical temperature for Bose-Einstein condensation 
for a given total atom number $N$ in the trap is here determined by \cite{Bag87},
\begin{equation}
 N=\int_0^1 d\epsilon g(\epsilon)
(\exp(\epsilon  U_0/k_{\rm B}T_{\rm C})-1)^{-1} \ .
\label{eqkrittemp}
\end{equation}

\begin{widetext}

\begin{figure}[htb!]
\includegraphics[width=130mm]{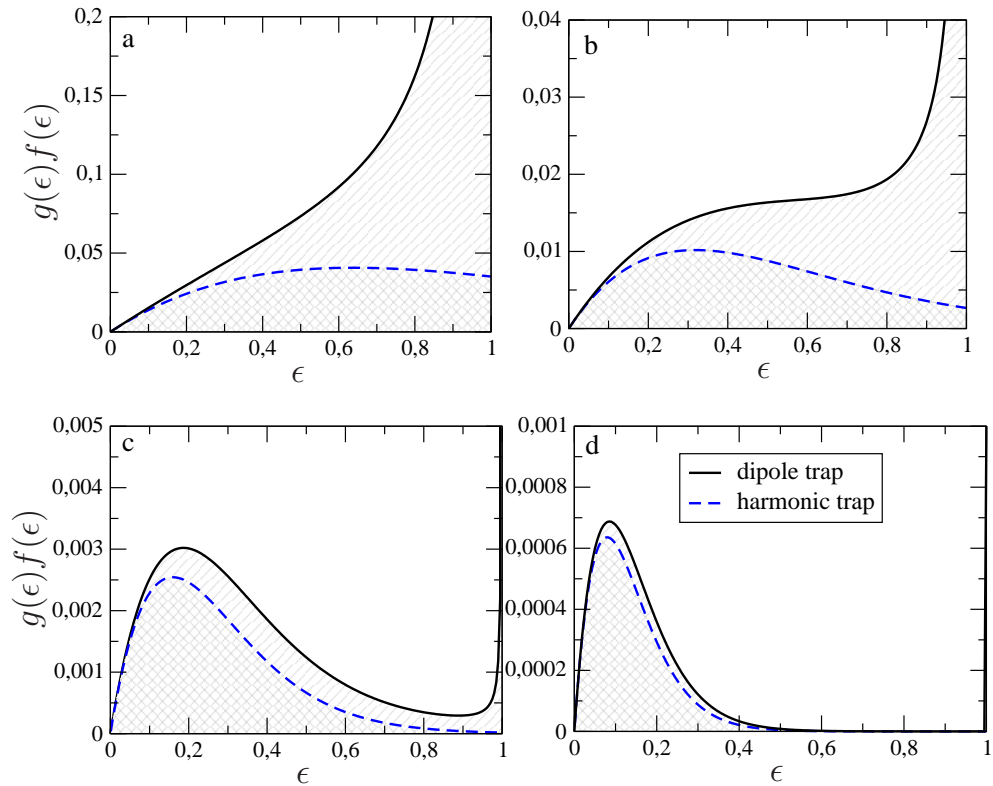}
\caption{Comparison of the number density $g(\epsilon)f(\epsilon)$ of the
dipole trap potential
(black, solid line) and its harmonic approximation (blue, dashed line)
for $\eta = U_0/k_{\rm B} T =  2.5, 5, 10, 20$.
The areas below the graphs must be finite to allow for Bose-Einstein
condensation.}
\label{figprodboszust}
\end{figure}
\end{widetext}

In case the integral diverges there is no Bose-Einstein condensation as 
for the well-known case of the one-dimensional harmonic oscillator or
the two-dimensional box potential.
Figure \ref{figprodboszust} shows the integrand $g(\epsilon)f(\epsilon)$ 
of expression (\ref{eqkrittemp}) with the Bose distribution
$f(\epsilon) =(\exp(\eta\epsilon)-1)^{-1}$
for different ratios $\eta=U_0/k_{\rm B}T$ of
trap depth to thermal energy.

\begin{figure}[htb]
\includegraphics[width=65mm]{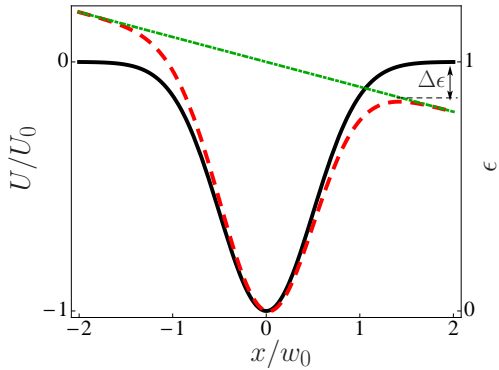}
\caption{The dipole trap potential without gravity (black, solid),
including gravity (red, dashed), and the pure gravitational potential
(green, dashed-dotted) at $y=z=0$. Due to gravity,
atoms may escape from the trap as soon as their energy is
$\epsilon > 1-\Delta \epsilon$. Thus, the real
potential is cut off at $\epsilon_{\rm t}=1-\Delta \epsilon$. }
\label{figpotential}
\end{figure}

The figure reveals that the number density $g(\epsilon)f(\epsilon)$ 
diverges for the dipole trap near the potential edge while it
stays finite for its harmonic approximation. Well below the singularity,
i.e. for small energies, the agreement between the two improves for
increasing $\eta$, i.e. for lower temperatures:
then the atoms are closer to the trap minimum, where the potential 
is well characterized by the harmonic approximation. For higher temperatures 
the atoms populate states of higher energy, where the harmonic potential 
differs significantly from the dipole trap. 
For $\epsilon \rightarrow 1$, the product $g(\epsilon)f(\epsilon)$ keeps 
the singularity from the density of states and it turns out that it is not
integrable. 

We have to conclude that -- strictly speaking -- the Bose Einstein phase 
transition does not occur in a single-beam optical dipole trap, no matter how 
cold the gas; which contradicts
findings of recent experiments \cite{Cen03}. 

In order to shed light on this curious result we have to look at the
true experimental conditions in more detail. 
A solution of this apparent contradiction can be
found in gravity -- which we neglected up to now. Including
gravity, the true potential reads
\begin{equation}
U_{\rm dip+g}(\vec r)=-\frac{U_0}{\left(1+z^2/z_0^2\right)}
\exp\left(\frac{-2(x^2+y^2)}{w_0^2\left(1+z^2/z_0^2\right)}\right)-mgx \ .
\end{equation}
As displayed in Figure \ref{figpotential}, gravity lowers the threshold 
energy $\epsilon_{\rm t}$ an atom needs to escape 
the trap in one spatial direction. Based on the assumption of sufficient 
ergodicity, we simply assume that gravity effectively leads to a cutoff
at a certain scaled energy $\epsilon_{\rm t}=1-\Delta \epsilon$.
The cutoff parameter
$\Delta \epsilon$ depends on 
the absolute trap depth $U_0$ and on the mass of the atoms. 

As a consequence of this cutoff, the integral over the number density
in equation (\ref{eqkrittemp}) extends up to 
$\epsilon_{\rm t}=1-\Delta \epsilon$ only. The
integral remains finite and Bose condensation may take place.

In fact, the critical temperature for a given number of atoms can then
well be estimated as usual within the harmonic approximation. To give an 
example, in an optical dipole trap with $U_0=200 {\rm \mu K}$, the critical 
temperature for 
$10^5$ $^{87}\mathrm{Rb}$-atoms is approximately $800 {\rm n K}$, which is in 
accordance with $\eta\approx 250$. This implies that the atoms are trapped
so deeply that the harmonic approximation is clearly valid --
see Figure \ref{figpotentiale}. The same conclusion can be drawn from
Figures \ref{figprodboszust}: while for fairly high temperatures ($\eta=2.5, 5$)
the number density in the dipole trap differs from the harmonic approximation
over the whole range of energies, this seizes to be true for low temperatures
($\eta=20$): here the only difference is the sharp singularity near
threshold ($\epsilon \approx 1$) which, as we have argued, turns out to
be irrelevant due to gravity. 

\section{Evaporative Cooling}
\label{secevcooling}
In the last section we came to the conclusion that even though the 
integral (\ref{eqkrittemp}) is not integrable in the usual limits, 
a critical temperature for Bose-Einstein condensation exists due to gravity
or any other interaction that effectively cuts off or blurs the
singularity as $\epsilon \rightarrow 1$.
The harmonic approximation of the potential is sufficient to determine $T_c$
because temperatures are typically low enough. However, for processes
involving higher temperatures, as for instance evaporation dynamics,
the number density $g(\epsilon)f(\epsilon)$ in the dipole trap 
differs significantly from the harmonic one and deviations from the
harmonic approximation might become relevant.

Evaporative cooling in optical traps is a simple and efficient method for 
reaching low enough temperatures to produce degenerate Bose \cite{Bar01} and 
Fermi \cite{Gran02} gases. The technique is based on the preferential 
removal of atoms with higher energy than the average energy. Subsequently, 
the remaining atoms thermalize through elastic collisions, which leads to a 
lower average energy and temperature. Our discussion here follows
\cite{Wal96} which describes evaporation in a classical approximation.
The latter is justified as typical temperatures are still much
higher than the quantum level 
spacing of an atom in the trap potential.
On the other hand the temperatures 
shall be low enough to work in the s-wave scattering regime with an energy 
independent cross section $\sigma=8\pi a^2$, where $a$ is the scattering 
length.

A central ambition is to determine the evolution of the energy distribution 
of a trapped gas during so-called plain evaporation \cite{Wal96}. 
Quite generally, well above the transition temperature, the trapped 
gas is well described by a classical phase-space distribution 
$f({\bf r},{\bf p})$, which is normalized according to
\begin{equation}
N=\frac{1}{(2\pi\hbar)^3}\int d^3rd^3p \ f({\bf r},{\bf p}) \ .
\end{equation}
The time evolution of $f({\bf r},{\bf p})$ is determined by the
Boltzmann equation \cite{Hua63}
\begin{equation}
\left(\frac{{\bf p}}{m} \nabla_{\rm r} - \nabla_{\rm r} U \nabla_{\rm p} 
+\frac{\partial}{\partial t}\right)
f({\bf r},{\bf p})={\mathcal I}({\bf r},{\bf p}),
\label{eqboltzmann}
\end{equation}
with the collision integral ${\mathcal I}$ for s-wave scattering and 
thus an energy-independent cross section $\sigma$ given by
\begin{widetext}
\begin{equation}
{\mathcal I}({\bf r},{\bf p_4})=\frac{\sigma}{2\pi m(2\pi\hbar)^3}
\int d^3p d\Omega' q\{f({\bf r,p_1})f({\bf r,p_2})-
f({\bf r,p_3})f({\bf r,p_4})\} \ .
\end{equation}
\end{widetext}
As in \cite{Wal96}, ${\bf p_3}$ and ${\bf p_4}$ are the momenta of two atoms 
before the collision, ${\bf p_1}={\bf P}/2+{\bf q'}$ and 
${\bf p_2}={\bf P}/2-{\bf q'}$ the momenta after the collision. Here we 
denote a relative momentum with ${\bf q}=({\bf p_3}-{\bf p_4})/2$, the total
momentum with ${\bf P}={\bf p_3}+{\bf p_4}$, and $|{\bf q'}|=|{\bf q}|$. 
$\Omega'$ denotes the direction of ${\bf q'}$ with respect to ${\bf q}$. \\
The assumption of sufficient ergodicity leads to a phase space distribution 
as a function of the scaled single-particle energy $\epsilon$ only, such 
that it can be written as
\begin{equation}
f({\bf r},{\bf p})= N \int d\epsilon \  
\delta \left((U({\bf r})+p^2/2m)/U_0-\epsilon  \right)f(\epsilon) \ ,
\label{eqphasespacedistr}
\end{equation}
where $f(\epsilon)$ is the distribution of atoms as a function of scaled
energy $\epsilon$ which we define to be normalized to one. 
The substitution (\ref{eqphasespacedistr}) leads to a rigorous simplification 
of the Boltzmann equation (\ref{eqboltzmann}). After applying 
$(2\pi \hbar)^{-3} \int d^3r d^3p \ \delta \left(U({\bf r})+p^2/2m-\epsilon 
\right)$ to both sides of (\ref{eqboltzmann}), the gradient terms 
on the left-hand side sum to zero, and only a term
$g(\epsilon)\partial f(\epsilon) /\partial t$ survives.
The right hand side can also be simplified, as described in \cite{Wal96}.
As a general result, the following kinetic equation for the time evolution of 
the distribution function $f(\epsilon)$ 
(initially normalized to one) of a gas in a trap with 
density of states $g(\epsilon)$ is obtained:

\begin{widetext}
 \begin{equation}
  g(\epsilon_4)\dot f (\epsilon_4)=
\int d\epsilon_1 d\epsilon_2 d\epsilon_3 
\delta(\epsilon_1+\epsilon_2-\epsilon_3-\epsilon_4)
g(\min(\epsilon_1, \epsilon_2,\epsilon_3, \epsilon_4))
\{f(\epsilon_1)f(\epsilon_2)-f(\epsilon_3)f(\epsilon_4)\} \ .
\label{eqkineticeq}
\end{equation}
Here, dot denotes time derivative 
$\dot f =\frac{d f} {d \tau}$ with respect to a dimensionless
time $\tau = N_0 \frac{m \sigma U_0^2} {\pi^2 \hbar^3} t$ where
$N_0$ is the initial number of atoms. Recall that $f(\epsilon)$
is normalized to one initially. Its
time evolution is obtained
numerically by writing the integral as a sum over a discretized energy scale.
In Figure \ref{fignumericplain} we show evaporation dynamics 
for dipole trap (top) and for the truncated
harmonic potential (bottom).
Furthermore, the evolution of the number density $g(\epsilon)f(\epsilon)$ 
is shown for the two potentials; its integral reflects the number of atoms 
remaining in the trap. 

\begin{figure*}[htb!]
\includegraphics[width=140mm]{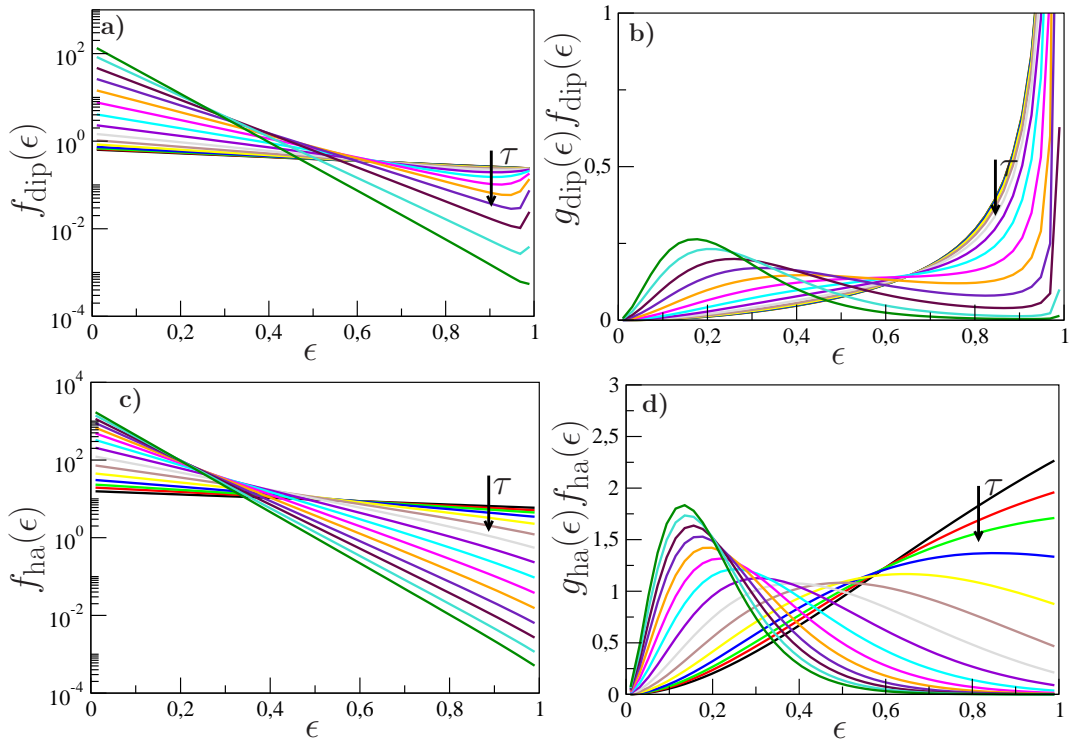}
\caption{Time evolution of normalized distribution function 
$f(\epsilon)$  during
plain evaporative cooling in a dipole trap (a) and a harmonic trap (c);
time evolution of the number density $g(\epsilon)f(\epsilon)$ for the
dipole (b) and the harmonic trap (d). Initially, $f(\epsilon)$ is
chosen to correspond to a Boltzmann distribution with $k_{\rm B} T = U_0$,
normalized to one.  The arrows indicate
increasing time; time steps grow exponentially.}
\label{fignumericplain}
\end{figure*}
\end{widetext}

We see from Figure \ref{fignumericplain} that for all times, $f(\epsilon)$ can 
well be approximated by an exponential, i.e. a Boltzmann distribution. 
Apparently, during evaporation the thermal nature of the distribution is 
preserved. In the case of the dipole trap potential, it is only at the 
edge of the 
potential well that $f(\epsilon)$ differs somewhat from an exponential, due to 
the singularity in the density of states. 

Being a Boltzmann distribution, 
the slope of the graphs allows us to extract a temperature of the 
atoms remaining in the trap.

Furthermore, as the graphs in Figure \ref{fignumericplain} a) and c) 
reveal, the rate of change of 
$f(\epsilon)$ for the dipole trap is smaller than for the harmonic potential.
The same conclusion can be drawn from parts $b)$ and $d)$ of the same 
figure. As the integral over $g(\epsilon)f(\epsilon)$ equals the number of 
trapped atoms, one finds that the evaporation process in the truncated 
harmonic trap happens much faster than in the dipole trap. However, 
at the end of the evaporation less atoms are left in the dipole trap, 
while the final temperature is approximately the same. 

In order to investigate these observations in more detail, we calculate
the number of atoms remaining in the trap by integrating over  
the number density $g(\epsilon)f(\epsilon)$. Moreover, we determine the 
temperature of the remaining atoms by fitting a Boltzmann distribution 
to $f(\epsilon)$, as a function of time. 

Before we discuss these results, however, we want to turn to the semianalytical 
solutions of equation (\ref{eqkineticeq}), which are described in \cite{Wal96}. 
The starting point is the replacement of the exact energy distribution 
function by a truncated Boltzmann distribution, which we saw is pretty well
justified by looking at Figure \ref{fignumericplain};
\begin{equation}
 f(\epsilon,t)\approx a(t) \exp(-U_0 \epsilon /k_{\rm B} T(t))
\Theta(\epsilon_{\rm t}-\epsilon) \ .
\label{eqtruncboltz}
\end{equation}
The Heaviside step function ensures that $f(\epsilon,t) = 0$ for 
$\epsilon > \epsilon_{\rm t}$; particles with an energy above threshold 
escape from the trap. 
It is worth mentioning that $T=T(t)$ is not a temperature in the strict 
thermodynamic sense, since evaporation is a nonequilibrium process; 
but it surely is a convenient parameter characterizing the essentially 
nonequilibrium distribution.
For the sake of clarity we introduce the dimensionless ``temperature'' 
$\tilde{T}=k_{\rm B}T/U_0$ and a scaled total internal 
energy $\tilde{E}=E/(U_0 N_0)$. 
We keep the single-particle energy $\epsilon$ scaled to $U_0$ as well. 

\begin{figure}[htb]
\includegraphics[width=80mm]{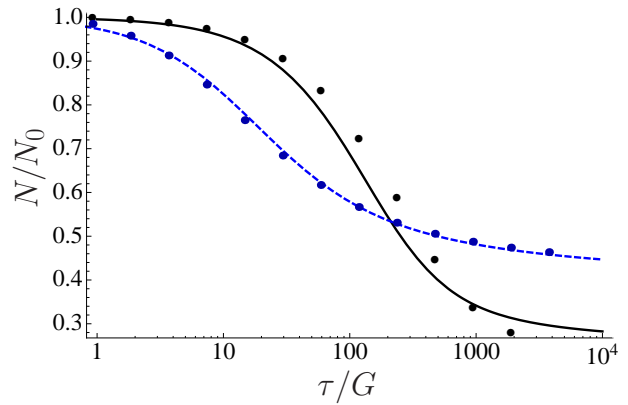}
\caption{Fraction of atoms $N/N_0$ remaining in the trap as a function 
of scaled time $\tau/G$ after starting evaporation from a scaled 
temperature $\tilde{T}=k_{\rm B}T/U_0=1$. The black, solid line is the result for the 
dipole trap and the blue dashed for its harmonic approximation 
obtained by integration of the differential equations resulting 
from the truncated Boltzmann approximation. The squares and circles 
are the results from the fully numerical treatment for dipole and 
harmonic trap, respectively.}
\label{figevapN}
\end{figure}
\begin{figure}[htb]
\includegraphics[width=80mm]{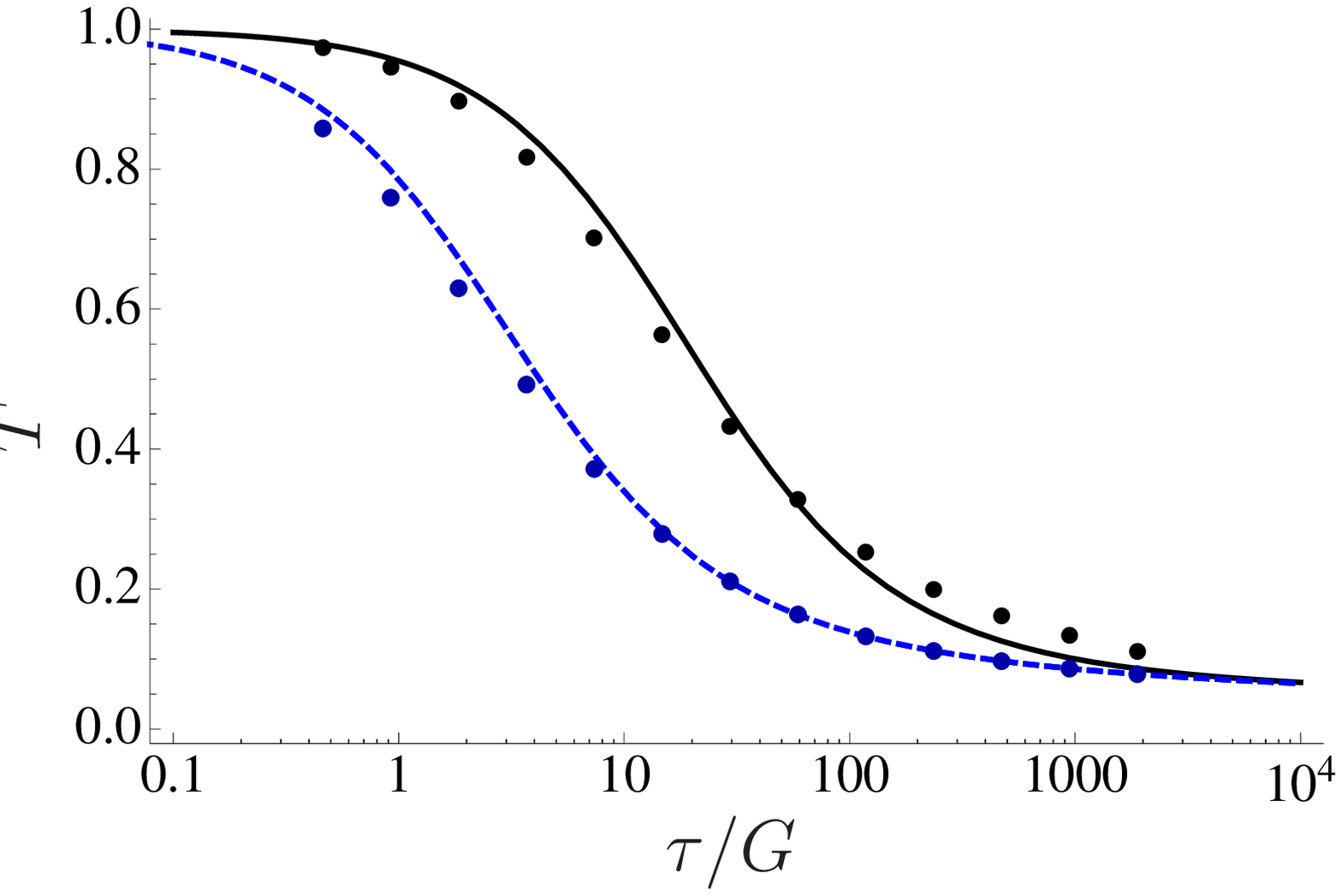}
\caption{Scaled temperature  $\tilde{T}=k_{\rm B}T/U_0$ of the remaining 
atoms as a 
function of scaled time $\tau/G$ after starting evaporation from a 
temperature $\tilde{T}=1$. The black, solid line is the result for the 
dipole trap and the blue dashed line for its harmonic approximation 
obtained by integration of the differential equations resulting from 
the truncated Boltzmann approximation. The squares and circles result 
from the fully numerical treatment for dipole and harmonic trap, 
respectively.}
\label{figevapT}
\end{figure}

Substituting the truncated Boltzmann distribution (\ref{eqtruncboltz}) in 
equation (\ref{eqkineticeq}), the integration can be performed easily. 
In order to determine the rate of change of the number of trapped atoms 
one integrates over untrapped energy states 
with $\epsilon_4 = \epsilon_1+\epsilon_2-\epsilon_3 > \epsilon_{\rm t} > 
\epsilon_1, \epsilon_2 > \epsilon_3$: 
\begin{eqnarray}
 {\dot{N}_{\rm ev}/N_0}=-\int _{\epsilon_{\rm t}}^{\infty}d\epsilon_4 
g(\epsilon_4)\dot{f}(\epsilon_4)
\newline
=-a^2 
\tilde{T} e^{-\eta} V
\label{eqevap1}
\end{eqnarray}
with 
\begin{eqnarray}
 V=\int_0^{\epsilon_{\rm t}}d\epsilon 
g(\epsilon) \{(\epsilon_{\rm t}-\epsilon- 
\tilde{T})e^{-\epsilon/\tilde{T}}+\tilde{T}e^{-\eta}\} \ ,
\end{eqnarray}
which corresponds to the ``effective volume for elastic
collisions'' of \cite{Wal96}. Note that the factor
$a=a(\tau)$ in equ. (\ref{eqevap1}) has to be determined from
the normalization condition
$a(\tau) = (N(\tau)/N_0)/\int_0^{\epsilon_{\rm t}}d\epsilon g(\epsilon) 
e^{-\epsilon/\tilde{T}(\tau)}$.

The evaporation of atoms leads to a loss of internal energy of the 
trapped gas. Following the elaborations in \cite{Wal96}, we find
an evolution equation somewhat similar to (\ref{eqevap1}) for
the scaled total energy
 ${\tilde{E}}$.
On the other hand, the rate of change of the internal energy due to the 
evaporation of atoms is
\begin{equation}
 \dot{\tilde{E}}=C\dot{\tilde{T}}+\mu(\dot{N}/N_0)
\label{eqevap3}
\end{equation}
where $C$ denotes the heat capacity 
$C=(\partial \tilde{E}/\partial\tilde{T})_N$ and 
$\mu=(\partial \tilde{E}/\partial (N/N_0))_{\tilde{T}}=\tilde{E}/(N/N_0)$ 
a chemical potential. We thus arrive at
coupled differential equations which describe the rate of change of 
temperature, atom number and internal energy during evaporation in 
closed form. These can be solved numerically.

Using this approximative set of differential equations,
we want to investigate the evaporation process in the dipole trap
based on the true density of states (\ref{eqzustandsdichtedipolfalle}) and
compare with results for the harmonic approximation.
Furthermore, we compare with a full numerical treatment based
on the Boltzmann equation (\ref{eqkineticeq}).
In Figure \ref{figevapN} we show the evolution of the number of 
remaining atoms and in Fig. \ref{figevapT}
their temperature, respectively, during evaporation. We choose
an initial scaled temperature of $\tilde{T}=k_{\rm B}T/U_0=1$. 
Recall that time is dimensionless with
$\tau = N_0 \frac{m \sigma U_0^2} {\pi^2 \hbar^3} t$ with
$N_0$ the initial number of atoms in the trap
(similar to \cite{Wal96}, times are scaled to $\tau/G$).
The curves are obtained 
by solving the above mentioned coupled differential equations based on 
the truncated Boltzmann approximation, while the symbols follow from the 
numerical solution of equ. (\ref{eqkineticeq}),
by integrating over $g(\epsilon)f(\epsilon)$ for 
the number of atoms and by fitting an exponential to $f(\epsilon)$ 
to determine the temperature.

Figures \ref{figevapN} and \ref{figevapT} confirm that the evaporation process 
in the dipole trap is indeed much slower than in a harmonic trap
as already indicated in Figure \ref{fignumericplain}. 
The harmonic approximation may lead to predictions
that differ by more than an order of magnitude.
Furthermore, we see that in a dipole trap, more atoms get lost
eventually compared to the harmonic trap. The final temperature,
on the other hand, turns out to be pretty much the same. 

In the case of the dipole trap the truncated Boltzmann approximation does not 
describe the distribution function $f(\epsilon)$ accurately
over the whole energy range. Clearly, we
see deviations for energies near the threshold in 
Figure \ref{fignumericplain}. These explain the 
differences between
the numerical and the approximative results
for atom number and temperature in
Figures \ref{figevapN} and \ref{figevapT}.

\section{A Bose-Einstein condensate in a single-beam dipole trap}
\label{secbec}
\begin{figure}[htb]
\includegraphics[width=75mm]{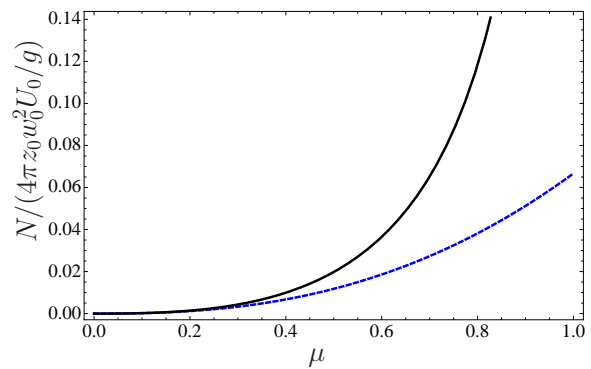}
\caption{Chemical potential versus atom number of a condensate in a dipole trap
(black, solid) and a harmonic trap (blue, dashed)
in Thomas-Fermi approximation.}
\label{figchempottf}
\end{figure}

We here assume that a Bose-Einstein
condensate has been created in or transferred to a 
single-beam optical dipole trap. At temperatures well below
the critical temperature, the condensate wave function is well 
described by the mean field Gross-Pitaevskii equation
\begin{equation}
\left(-\frac{\hbar^2}{2m}{\bf \Delta} + U(\bf r)\right)\psi(\bf r)
+g|\psi(\bf r)|^2\psi(\bf r) = \mu\psi(\bf r)
\end{equation}
with $\mu$ the chemical potential.
Often, the so-called Thomas-Fermi approximation may be employed
\cite{Pet02, Pit03}
which describes the BEC in the limit of large atom numbers such that
the contribution of the kinetic energy may be neglected with
respect to the internal energy and the potential energy.
The condensate wave function is then given by the simple formula
\begin{equation}
|\psi(\bf r)|^2 = \frac{\mu-U(\bf r)}{g},
\label{eqThomas-Fermi}
\end{equation}
valid for all positions ${\bf r}$ such that the right hand side is positive,
and $\psi(\bf r)=0$ otherwise.
The parameter $g$ describes the strength of the atom-atom interaction and 
is connected to the s-wave scattering length $a$ through 
$g=4 \pi \hbar ^2 a/m$. For a harmonic approximation, 
the Thomas-Fermi wave function (\ref{eqThomas-Fermi})
leads to the analytical expression $\mu=(15 g N / 4 \pi w_0^2 z_0 U_0)^{2/5}$ 
for the chemical potential,
where $\mu$ is normalized to $U_0$ and positive (similar to $\epsilon$). 
Remarkably, in Thomas-Fermi approximation, even the full dipole
potential (\ref{eqdipolpot}) allows for an analytical treatment. 
By integrating expression (\ref{eqThomas-Fermi}) over ${\bf r}$ and equating 
it with the total atom number $N$ we find the simple expression
\begin{eqnarray}
 N(\mu) &=& \frac{4 \pi}{9 g}z_0w_0^2 U_0 \Bigg(\sqrt{\frac{\mu}{1-\mu}} 
(4 \mu -3)
\nonumber
\\
&-& 3(\mu -1)\arctan\left(\sqrt{\frac{\mu}{1-\mu}}\right)\Bigg) \ .
\end{eqnarray}
Similar to the density of states (\ref{eqzustandsdichtedipolfalle}), 
the number of particles $N$ 
as a function of the chemical potential $\mu$ develops a singularity at 
$\mu=1$ in the dipole trap. Clearly, in the harmonic trap the particle number
for any $\mu$ remains finite. For small condensate numbers $N$ the chemical 
potentials agree, as shown in Figure \ref{figchempottf}. 

Furthermore, we found an analytical expression for the internal (interaction) energy in 
Thomas-Fermi approximation for a condensate. 
By integrating the squared expression (\ref{eqThomas-Fermi}),
one finds for the harmonic approximation the well-known 
$E_{\rm int}(N)=\frac{2}{7}U_0
\left(\frac{15 g}{4 \pi z_0 w_0^2 U_0}\right)^{2/5} N^{7/5}$, while for
the dipole trap the result can only be given analytically as a function of 
the chemical potential:
\begin{eqnarray}
E_{\rm int}(\mu) &=&  \frac{\pi z_0 w_0^2 U_0^2}{4 g}
\bigg(\sqrt{\frac{\mu}{1-\mu}}\left(\frac{23}{9}(\mu-1)
+\frac{38}{9}(\mu-1)^2\right)
\nonumber
\\ 
 &+&  (1-\frac{8}{3}(\mu-1)^2)\arctan
\left(\sqrt{\frac{\mu}{1-\mu}}\right)\bigg).
\end{eqnarray}
Clearly, the internal energy remains finite
for any finite number of atoms. 

\begin{figure}[htb]
\includegraphics[width=75mm]{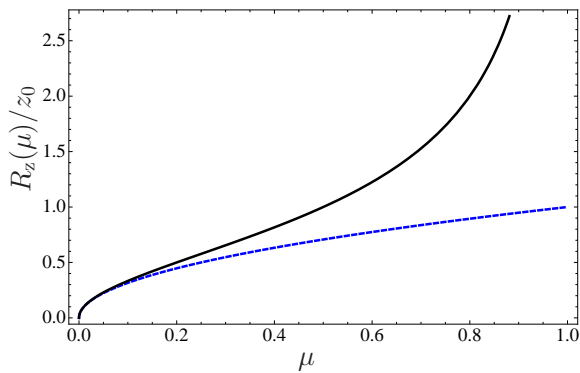}
\caption{Thomas-Fermi radius along the propagation direction of the laser 
versus chemical potential for dipole trap (black, solid) and its harmonic 
approximation (blue, dashed).}
\label{figtfradiusztf}
\end{figure}

Finally, we determine the Thomas-Fermi radii of the condensate in such an 
optical dipole trap. They are defined by means of the condition 
$U({\bf r})=\mu$ and characterize the extensions of the condensate
in the various directions.
For the harmonic potential they are well known to be 
$R_{\rm \rho}(\mu)=\frac{w_0}{\sqrt{2}} \sqrt{\mu}$ and 
$R_{\rm z}(\mu)=z_0\sqrt{\mu}$. For the dipole trap, we find easily
\begin{eqnarray}
 R_{\rm \rho}(\mu)=\frac{w_0}{\sqrt{2}}\sqrt{-\ln(1-\mu)} \ \ ;
  \ \ R_{\rm z}(\mu)=z_0 \sqrt{\frac{\mu}{1-\mu}} \ .
\end{eqnarray}
Clearly, as $\mu\rightarrow 0$, these expressions coincide with the
harmonic case. In general, however, in particular along the z-axis
(direction of propagation of the laser), the Thomas-Fermi radius in 
the dipole trap differs strongly from the one obtained in the harmonic 
trap for large $\mu$. Both expressions diverge for $\mu\rightarrow 1$.
In Figure \ref{figtfradiusztf} we show $R_{\rm z}(\mu)$ for 
both traps. 

It should be emphasized that temperatures and atom numbers 
that may be achieved with evaporative cooling as described in the first
part of this paper lead to ``small'' condensates that may well be described
with the harmonic approximation. Nevertheless, it is conceivable
that much larger condensates are created in some other trap before
being loaded into a single-beam optical dipole trap such that
deviations from the harmonic trap become relevant and our results apply.

\section{Conclusion}\label{secconclusion}
We found an analytical expression for the density of states of an
optical dipole trap potential, consisting of one single focused 
laser beam. The density of states has a singularity at the 
potential edge of the trap. 

Without further consideration, this singularity would prevent Bose-Einstein
condensation from occurring. However, since an effective cutoff will be present
in current-day experiments (e.g. due to gravity), condensation can be
achieved. Then, near condensation, the usual harmonic approximation
is valid and may be used to determine, e.g., 
the critical temperature for Bose-Einstein condensation.

Still, the existence of this singularity makes itself felt whenever
states of higher energy are involved as, for instance, in evaporative
cooling. We found that plain evaporative cooling is much slower 
(about one order of magnitude) in
the dipole trap compared to its harmonic approximation.
In addition, for the optical dipole trap more atoms get 
lost on the way to the same temperature compared to the harmonic trap. 

Finally, we determine analytical expressions for chemical potential, 
internal energy and Thomas-Fermi radii for the true optical dipole trap 
in Thomas-Fermi approximation. It turns out that the chemical potential 
and the radii of the condensate develop singularities 
similar to the density of states. 

While we here concentrate on bosonic gases, it will be interesting
to see how the semiclassical 
expression (\ref{eqzustandsdichtedipolfalle}) will be of 
relevance for the behavior of fermionic gases in single-beam 
dipole traps. 

\section*{Acknowledgments}
This work came into being due to many fruitful discussions we shared
with Hanspeter Helm and his co-workers in Freiburg, in particular 
Christoph K\"afer.
L. S. acknowledges support from the International Max Planck Research
School (IMPRS), Dresden.

\end{document}